\documentclass[12pt]{article}
\usepackage{graphicx}
\begin{document}
\begin{titlepage}

\title{The Pound-Rebka experiment and torsion in the
Schwarzschild spacetime}

\author{J. W. Maluf$\,^{(a)}$, S. C. Ulhoa$\,^{(b)}$ and
F. F. Faria$\,^{(c)}$\\
Instituto de F\'{\i}sica, \\
Universidade de Bras\'{\i}lia\\
C. P. 04385 \\
70.919-970 Bras\'{\i}lia DF, Brazil\\}
\date{}
\maketitle
\begin{abstract}
We develop some ideas discussed by E. Schucking [arXiv:0803.4128]
concerning the geometry of the gravitational field. First, we address
the concept according to which the gravitational acceleration is a 
manifestation of the spacetime torsion, not of the curvature tensor.
It is possible to show that there are situations in which the 
geodesic acceleration of a particle may acquire arbitrary values, 
whereas the curvature tensor approaches zero. We conclude that the
spacetime curvature does not affect the geodesic acceleration. 
Then we consider the Pound-Rebka experiment, which
relates the time interval $\Delta \tau_1$ of two light 
signals emitted at a position $r_1$, to the time interval 
$\Delta \tau_2$ of the signals received at a position $r_2$, in a 
Schwarzschild type gravitational field. The experiment is determined
by four spacetime events. The infinitesimal vectors formed by these
events 
do not form a parallelogram in the (t,r) plane. The failure in the
closure of the parallelogram implies that the spacetime has torsion.
We find the explicit form of the torsion tensor that explains the 
nonclosure of the parallelogram. 
\end{abstract}
\thispagestyle{empty}
\vfill
\noindent PACS numbers: 04.20.Cv, 04.20.-q, 04.80.Cc\par
\bigskip
\noindent (a) wadih@unb.br\par
\noindent (b) sc.ulhoa@gmail.com\par
\noindent (c) fff@unb.br\par
\end{titlepage}
\newpage

\section{Introduction and notation}
The spacetime geometry is determined by the metric tensor
$g_{\mu\nu}$, and the dynamics of the metric tensor is determined by
Einstein's equations. For a given metric tensor there exists an
infinity of tetrad fields $e^a\,_\mu$
that are compatible with the spacetime geometry. Tetrad fields
may be interpreted as reference frames adapted to a class of 
observers in spacetime. Einstein's equations may be written in the
traditional form in terms of the metric tensor, in which case the 
curvature tensor plays a prominent role, or in terms of the tetrad 
field. In the latter case the field equations are constructed out of
the torsion tensor. Therefore the dynamics of the gravitational field
admits a description either in terms of the curvature tensor (of the
Levi-Civita connection), or of the torsion tensor (of the 
Weitzenb\"ock  connection) \cite{Hehl1}.

However, there is a point of view according to which
the gravitational force that acts on a particle or on a frame, in a 
given gravitational field, is due to the torsion tensor only, not
to the spacetime curvature. Of course the curvature tensor is 
responsible for the tidal forces, but the force on a particle that 
moves along a particular geodesic
worldline $x^\mu(s)$, with tangent vector 
$u^\mu=dx^\mu/ds$, is due to the torsion tensor. This is one of the
issues discussed by Schucking \cite{Schucking}, and we will address
it in this paper, in some detail, in terms of the acceleration 
tensor. This tensor is a coordinate invariant quantity that describes
the accelerations that are necessary to maintain a reference frame in
spacetime in a given inertial state (for instance, to maintain the
frame in stationary state). The reference frame is fixed by
identifying the timelike components of the inverse tetrad field with
the velocity field of the class of observers, i.e., 
$e_{(0)}\,^\mu = u^\mu$.

A second issue to be considered here is the interpretation
of the Pound-Rebka experiment as a manifestation of the spacetime
torsion, as suggested by Schucking \cite{Schucking}. 
Suppose that at the top of a tower, at a distance $r_1$ from
the center of the Earth a light signal is emitted radially downwards 
at the instant $t_1$, and received at a position $r_2$ at the instant
$t_2$. After a proper time interval $\Delta \tau_1$ a second light 
signal is emitted downwards, and is received at the position $r_2$ at 
the instant $t_2+\Delta \tau_2$. It is known that timelike and null 
vectors formed by the events 
$(t_1, r_1)$, $(t_1 + \Delta \tau_1, r_1)$, $(t_2, r_2)$ and 
$(t_2 + \Delta \tau_2, r_2)$ do not form a parallelogram in the (t,r)
plane. The nonclosure of the parallelogram may be interpreted as a
manifestation of the torsion of the spacetime. By establishing the
frame of stationary observers in the Schwarzschild spacetime we 
arrive at the torsion tensor that precisely explains the breaking of
the parallelogram. This issue will be investigated in detail in the
present analysis.

The paper is organized as follows. In section 2 we review the 
construction of the acceleration tensor. The values of this tensor
characterize the 
inertial state of the frame, i.e., it provides the nongravitational
accelerations (translational and angular velocity of the local 
spatial frame with respect to a nonrotating Fermi-Walker transported
frame) that are exerted on the frame. In section 3 we
discuss the possibility of having a situation in which the curvature
tensor approaches zero, whereas the geodesic (gravitational)
acceleration of a particle may acquire arbitrary values. The geodesic
acceleration is related to some components of the acceleration
tensor (constructed out of the torsion tensor) for a stationary frame
in spacetime.
In section 4 we consider the Pound-Rebka experiment and explain the
breaking of the parallelogram in terms of the spacetime torsion.
We conclude that the torsion tensor is an important entity in the
description of the spacetime geometry and of the gravitational 
field.\par
\bigskip

Notation: space-time indices $\mu, \nu, ...$ and SO(3,1) indices
$a, b, ...$ run from 0 to 3. Time and space indices are indicated
according to $\mu=0,i,\;\;a=(0),(i)$. The tetrad field is denoted
$e^a\,_\mu$, and the torsion tensor reads
$T_{a\mu\nu}=\partial_\mu e_{a\nu}-\partial_\nu e_{a\mu}$.
The flat, Minkowski space-time metric tensor raises and lowers
tetrad indices and is fixed by
$\eta_{ab}=e_{a\mu} e_{b\nu}g^{\mu\nu}=diag(-1,1,1,1)$. The 
determinant of the tetrad field is represented by 
$e=\det(e^a\,_\mu)$.

The torsion tensor defined above is often related to the object of
anholonomity $\Omega^\lambda\,_{\mu\nu}$ via 
$\Omega^\lambda\,_{\mu\nu}= e_a\,^\lambda T^a\,_{\mu\nu}$.
However, we assume that the spacetime geometry is defined by the 
tetrad field only, and in this case the only
possible nontrivial definition for the torsion tensor is given by
$T^a\,_{\mu\nu}$. This torsion tensor is related to the 
antisymmetric part of the Weitzenb\"ock  connection 
$\Gamma^\lambda_{\mu\nu}=e^{a\lambda}\partial_\mu e_{a\nu}$, which
is frame dependent and 
establishes the Weitzenb\"ock spacetime. The curvature of the
Weitzenb\"ock connection vanishes. However, the tetrad
field also yields the metric tensor, which establishes the
Riemannian geometry. Therefore in the framework of 
a geometrical theory based only on the tetrad field one may use the 
concepts of both Riemannian and Weitzenb\"ock geometries.

\section{The tetrad field as reference frame and the 
acceleration tensor}

We recall the discussion presented in
refs. \cite{Maluf1,Maluf2} regarding the characterization of tetrad 
fields as reference frames in spacetime. A frame may be characterized
in a coordinate invariant way by its inertial accelerations, 
represented by the acceleration tensor. 

We denote by $x^\mu(s)$ the worldline $C$ of an observer in 
spacetime, where $s$ is the proper time of the observer. The velocity
of the observer on $C$ reads $u^\mu=dx^\mu/ds$. We identify 
the observer's velocity with the $a=(0)$ component of $e_a\,^\mu$:
$u^\mu(s)=e_{(0)}\,^\mu$. The acceleration $a^\mu$ of the observer
is given by the 
absolute derivative of $u^\mu$ along $C$ \cite{Hehl},

\begin{equation}
a^\mu= {{Du^\mu}\over{ds}} ={{De_{(0)}\,^\mu}\over {ds}} =
u^\alpha \nabla_\alpha e_{(0)}\,^\mu\,, 
\label{1}
\end{equation}
where the covariant derivative
is constructed out of the Christoffel symbols. Thus $e_a\,^\mu$ and
its derivatives determine the velocity and acceleration along the
worldline of an observer. The set of tetrad fields for which 
$e_{(0)}\,^\mu$ describe a 
congruence of timelike curves is adapted to a class of 
observers characterized by the velocity field 
$u^\mu=e_{(0)}\,^\mu$ and by the acceleration $a^\mu$. 

We may consider not only the acceleration of observers along
trajectories whose tangent vectors are given by $e_{(0)}\,^\mu$, but
the acceleration of the whole frame along $C$. The 
acceleration of the frame is determined by the absolute derivative
of $e_a\,^\mu$ along the path $x^\mu(s)$. Thus, assuming that the 
observer carries an orthonormal tetrad frame $e_a\,^\mu$, the 
acceleration of the latter along the path is given by 
\cite{Mashh}

\begin{equation}
{{D e_a\,^\mu} \over {ds}}=\phi_a\,^b\,e_b\,^\mu\,,
\label{2}
\end{equation}
where $\phi_{ab}$ is the antisymmetric acceleration tensor. 
According to ref. \cite{Mashh}, 
in analogy with the Faraday tensor we can identify
$\phi_{ab} \rightarrow ({\bf a}, {\bf \Omega})$, where 
${\bf a}$ is the translational acceleration ($\phi_{(0)(i)}=a_{(i)}$)
and ${\bf \Omega}$ is the angular velocity 
of the local spatial frame  with respect to a nonrotating
(Fermi-Walker transported) frame. 
It follows that

\begin{equation}
\phi_a\,^b= e^b\,_\mu {{D e_a\,^\mu} \over {ds}}=
e^b\,_\mu \,u^\lambda\nabla_\lambda e_a\,^\mu\,.
\label{3}
\end{equation}

Therefore given any set of tetrad fields for an arbitrary 
gravitational field configuration, its geometrical interpretation
may be obtained by suitably interpreting the velocity field 
$u^\mu=\,e_{(0)}\,^\mu$ and the acceleration tensor $\phi_{ab}$.
The acceleration vector $a^\mu$ defined by Eq. (1) 
may be projected on a frame in order to yield

\begin{equation}
a^b= e^b\,_\mu a^\mu=e^b\,_\mu u^\alpha \nabla_\alpha
e_{(0)}\,^\mu=\phi_{(0)}\,^b\,.
\label{4}
\end{equation}
Thus $a^\mu$ and $\phi_{(0)(i)}$ are not different 
accelerations of the frame. 

The acceleration $a^\mu$ given by Eq. (1) may be rewritten as

\begin{eqnarray}
a^\mu&=& u^\alpha \nabla_\alpha e_{(0)}\,^\mu 
=u^\alpha \nabla_\alpha u^\mu =
{{dx^\alpha}\over {ds}}\biggl(
{{\partial u^\mu}\over{\partial x^\alpha}}
+\,^0\Gamma^\mu_{\alpha\beta}u^\beta \biggr) \nonumber \\
&=&{{d^2 x^\mu}\over {ds^2}}+\,^0\Gamma^\mu_{\alpha\beta}
{{dx^\alpha}\over{ds}} {{dx^\beta}\over{ds}}\,,
\label{5}
\end{eqnarray}
where $^0\Gamma^\mu_{\alpha\beta}$ are the Christoffel symbols.
Thus if $u^\mu=e_{(0)}\,^\mu$ represents a geodesic
trajectory, then the frame is in free fall and 
$a^\mu=0=\phi_{(0)(i)}$. Therefore we conclude that nonvanishing
values of $\phi_{(0)(i)}$ represent inertial accelerations
of the frame.

Following ref. \cite{Maluf1}, we take into account the 
orthogonality of the tetrads and write Eq. (3) as
$\phi_a\,^b= -u^\lambda e_a\,^\mu \nabla_\lambda e^b\,_\mu$, 
where $\nabla_\lambda e^b\,_\mu=\partial_\lambda e^b\,_\mu-
\,^0\Gamma^\sigma_{\lambda \mu} e^b\,_\sigma$. Next we consider
the identity $\partial_\lambda e^b\,_\mu-
\,^0\Gamma^\sigma_{\lambda \mu} e^b\,_\sigma+\,\,
^0\omega_\lambda\,^b\,_c e^c\,_\mu=0$, where
$^0\omega_\lambda\,^b\,_c$ is the metric compatible 
Levi-Civita connection, and express $\phi_a\,^b$ according to

\begin{equation}
\phi_a\,^b=e_{(0)}\,^\mu(\,\,^0\omega_\mu\,^b\,_a)\,.
\label{6}
\end{equation}
Finally we take into account the identity 
$\,\,^0\omega_\mu\,^a\,_b= -K_\mu\,^a\,_b$, where $-K_\mu\,^a\,_b$ 
are the Ricci rotation coefficients defined by

\begin{equation}
K_{\mu ab}={1\over 2}e_a\,^\lambda e_b\,^\nu(T_{\lambda \mu\nu}+
T_{\nu\lambda\mu}+T_{\mu\lambda\nu})\,,
\label{7}
\end{equation}
and $T_{\lambda \mu\nu}=e^a\,_\lambda T_{a\mu\nu}$. 
After simple manipulations we arrive at

\begin{equation}
\phi_{ab}={1\over 2} \lbrack T_{(0)ab}+T_{a(0)b}-T_{b(0)a}
\rbrack\,.
\label{8}
\end{equation}

The expression above is not invariant under local SO(3,1) 
transformations, and for this reason the values of $\phi_{ab}$
may characterize the frame.
However, eq. (8) is invariant under coordinate transformations. We 
interpret $\phi_{ab}$ as the inertial accelerations of the frame 
along the trajectory $C$.

In ref. \cite{Maluf1} we applied definition (8) to the analysis
of two simple configurations of tetrad fields in the flat Minkowski
spacetime. We considered the frame adapted to linearly accelerated
observers, and to a stationary frame whose four-velocity is
$e_{(0)}\,^\mu=(1,0,0,0)$ and which rotates around the $z$ axis.
The components $\phi_{(0)(i)}$ and $\phi_{(i)(j)}$ yield the 
known values of the translational acceleration and of the 
angular velocity of the frame, respectively. As we will see in 
section 3, in suitable situations the values of $\phi_{(0)(i)}$
which are necessary to maintain a frame in stationary state exactly
cancel the geodesic acceleration exerted on a particle or observer
in spacetime.

\section{Stationary frame and geodesic acceleration in the 
Schwarzschild spacetime}

In order to obtain the radial geodesic acceleration of $\,$ a 
particle in the Schwarzschild spacetime, as discussed in ref. 
\cite{Schucking}, we will address a more general situation, namely,
we will obtain the inertial accelerations 
that are necessary to impart to a frame such
that it remains stationary in spacetime. The Schwarzschild spacetime
is described by the line element

\begin{equation}
ds^2=-\biggl(1-{{2m}\over r}\biggr)dt^2 +
\biggl(1-{{2m}\over r}\biggr)^{-1} dr^2+r^2d\theta^2+
r^2\,(\sin\theta)^2\, d\phi^2\,.
\label{9}
\end{equation}

A field of stationary observers in spacetime is characterized by a
vector field $u^\mu$ such that $u^\mu=(u^0,0,0,0)$, i.e., the spatial
components of $u^\mu$ vanish. Thus in the construction of the tetrad
field we require 

\begin{equation}
e_{(0)}\,^i= u^i=0\,.
\label{10}
\end{equation}
In view of the orthogonality 
of the tetrad components this condition implies $e^{(k)}\,_0=0$.
A simple form of $e_{a\mu}$ in $(t,r,\theta,\phi)$ coordinates that
satisfies this property and yields (9) is given by

\begin{equation}
e_{a\mu}=\pmatrix{-\beta&0&0&0\cr
0&\alpha\sin\theta\,\cos\phi&r\cos\theta\,\cos\phi
&-r\sin\theta\,\sin\phi\cr
0&\alpha\sin\theta\,\sin\phi&r\cos\theta\,\sin\phi
&r\sin\theta\,\cos\phi\cr
0&\alpha\cos\theta&-r\sin\theta&0\cr}\,,
\label{11}
\end{equation}
where

\begin{eqnarray}
\alpha&=& \biggl(1-{{2m}\over r}\biggr)^{-1/2} \nonumber  \\ 
\beta &=& \biggl(1-{{2m}\over r}\biggr)^{1/2}  \,. 
\label{12}
\end{eqnarray}
In (11) $a$ and $\mu$ label lines and rows, respectively. 
It is possible to show that in the asymptotic limit 
$r\rightarrow \infty$ the inverse tetrad components in $(t,x,y,z)$
coordinates satisfy 

\begin{eqnarray}
e_{(1)}\,^\mu (t,x,y,z) & \cong &(0,1,0,0)\,, \nonumber \\
e_{(2)}\,^\mu (t,x,y,z) & \cong &(0,0,1,0)\,, \nonumber \\
e_{(3)}\,^\mu (t,x,y,z) & \cong &(0,0,0,1)\,.
\label{13}
\end{eqnarray}

We proceed now to determine the acceleration tensor $\phi_{ab}$.
After a number of manipulations we find that (11) represents a
nonrotating frame, i.e., 

\begin{equation}
\phi_{(i)(j)}=0\,.
\label{14}
\end{equation}
Altogether, conditions (13) and (14) fix the orientation of the frame
in spacetime.

The translational acceleration, however, is nonvanishing. From  
definition (8) we find

\begin{equation}
\phi_{(0)(i)}=T_{(0)(0)(i)} = 
e_{(0)}\,^\mu e_{(i)}\,^\nu T_{(0)\mu\nu}\,.
\label{15}
\end{equation}
For $a=(0)$ the only nonvanishing component of $T_{a\mu\nu}$ is
$T_{(0)01}=\partial_1 \beta$. The equation above yields

\begin{equation}
\phi_{(0)(i)}= g^{00}g^{11}e_{(0)0} e_{(i)1} T_{(0)01}\,,
\label{16}
\end{equation}
from what follows

\begin{eqnarray}
\phi_{(0)(1)}&=&{m\over r^2}\biggl( 1-{{2m}\over r}\biggr)^{-1/2}
\sin\theta\,\cos\phi \,,\nonumber \\
\phi_{(0)(2)}&=&{m\over r^2}\biggl( 1-{{2m}\over r}\biggr)^{-1/2}
\sin\theta\,\sin\phi \,,\nonumber \\
\phi_{(0)(3)}&=&{m\over r^2}\biggl( 1-{{2m}\over r}\biggr)^{-1/2}
\cos\theta \,.
\label{17}
\end{eqnarray}
We define the acceleration

\begin{equation}
{\bf a}= \phi_{(0)(1)} \hat{{\bf x}}+\phi_{(0)(2)} \hat{{\bf y}}
+\phi_{(0)(3)} \hat{{\bf z}}\,,
\label{18}
\end{equation}
which may be written as

\begin{equation}
{\bf a}= {m\over r^2}\biggl( 1-{{2m}\over r}\biggr)^{-1/2}\,
\hat{\bf r}\,.
\label{19}
\end{equation}
 
Equation (19) represents the inertial acceleration necessary to 
maintain the frame in stationary state in spacetime. Therefore it 
exactly cancels the geodesic acceleration that is exerted on the
frame. In fact,

\begin{equation}
a= {m\over r^2}\biggl( 1-{{2m}\over r}\biggr)^{-1/2}
\label{20}
\end{equation}
is precisely the geodesic acceleration obtained in ref. 
\cite{Schucking} by means of Cartan's structural equations or, for
instance, in ref. \cite{Hartle} by taking the absolute derivative 
(according to eq. (1)) of the velocity of a body in free fall in the
Schwarzschild spacetime. We note, however, that eq. (15) (and
consequently (19)) is invariant under coordinate transformations. 

Now we analise a very interesting consequence of eq. (20), 
considering that the acceleration $a$ is kept constant. Equation 
(20) is a quadratic equation for the mass $m$, which can be written
as

\begin{equation}
m^2+2a^2r^3\,m-a^2 r^4=0\,.
\label{21}
\end{equation}
Solving this equation for $m$ we find

\begin{equation}
m=-a^2r^3+ar^2(1+a^2r^2)^{1/2}\,,
\label{22}
\end{equation}
which, after simple manipulations, leads to

\begin{equation}
{m\over r}=a^2r^2\biggl[ \sqrt{1+{1\over {(ar)^2}}} -1\biggr]\,.
\label{23}
\end{equation}
Keeping in mind that $a$ is assumed to be nonvanishing and constant,
we define the variable $x$ according to 

\begin{equation}
{1\over {ar}}= x\,,
\label{24}
\end{equation}
and therefore

\begin{equation}
{m\over r}={1\over x^2}\biggl[ \sqrt{1+x^2}-1 \biggr]\,.
\label{25}
\end{equation}
When $r\rightarrow \infty$, $x\rightarrow 0$ and consequently

\begin{equation}
{1\over x^2}\biggl[ \sqrt{1+x^2}-1 \biggr] \cong {1\over 2}
-{x^2\over 8} \equiv {1\over 2} -\epsilon\,,
\label{26}
\end{equation}
where $\epsilon <<1$,
in the limit $r \rightarrow \infty$. Thus in this limit we have

\begin{equation}
{m\over r} \cong {1\over 2} -\epsilon\,,
\label{27}
\end{equation}
which implies $r\cong 2m(1+2\epsilon)$.
However the component of the curvature tensor (of the Levi-Civita
connection) in the $(t,r)$ plane is given by \cite{Schucking}

\begin{equation}
R_{0101}= {{2m}\over r^3}\,.
\label{28}
\end{equation}
Therefore, in view of eq. (27), in the limit when both 
$r \rightarrow \infty$ and $m \rightarrow \infty$
the curvature tensor vanishes,

\begin{equation}
R_{0101}= {m\over r} {2\over r^2}\cong {1\over r^2} \cong 0\,.
\label{29}
\end{equation}

Thus we see that if (i) $r \rightarrow \infty$ and (ii) 
${m/r} \cong {1/2} -\epsilon$, then the curvature tensor 
approaches zero whereas the geodesic acceleration $a$ may acquire
arbitrary values. This is not a realistic physical situation,
but it proves that the curvature tensor is not responsible for the
geodesic acceleration given by (20). This is the argument presented
by Schucking \cite{Schucking}: there may exist gravitational field
configurations such that
the curvature tensor approaches zero, whereas the geodesic
acceleration may acquire arbitrary values. The action of gravity on
a particle that undergoes geodesic acceleration is not affected by
the vanishing value of the curvature tensor. As we have seen, the
geodesic acceleration may be obtained from the acceleration tensor
given by eq. (8), and the latter 
is constructed out of the torsion tensor, which is ultimately 
responsible for the geodesic acceleration. We note that it
is impossible to write eq. (8) in terms of the curvature tensor.

\section{The Pound-Rebka experiment}

The relevance of the torsion tensor to the spacetime geometry 
is revealed by the Pound-Rebka experiment \cite{Pound}.
Let us consider the emission of two radial light signals at the 
position $r+\Delta r$ to the position $r$, in the Schwarzschild
spacetime. At the position $r+\Delta r$ the time elapsed between the
first and second signals is the proper time $d \tau_1$, and at $r$
the second signal is received after a proper time $d \tau_2$. As in 
the previous section, we assume that the Schwarzschild spacetime is
described the coordinates $(t,r,\theta,\phi)$. In this case we have

\begin{eqnarray}
d \tau_1&=& \beta(r+\Delta r) dt\, , \nonumber \\
d \tau_2&=& \beta (r) dt \,,
\label{30}
\end{eqnarray}
where $\beta(r)=(-g_{00})^{1/2}$. If $\Delta r / r <<1$, then

\begin{equation}
d\tau_1 \cong \biggl[ 1+ {{\Delta r}\over r^2}\biggl( {{GM}\over c^2}
\biggr) \biggr] d\tau_2\,,
\label{31}
\end{equation}
where we have used $m=GM/c^2$. The experimental verification of eq.
(31) is the result of the Pound-Rebka experiment \cite{Weinberg},
which may be described in Figure 1.

\begin{figure}[h]
\centering
\includegraphics[width=0.50\textwidth]{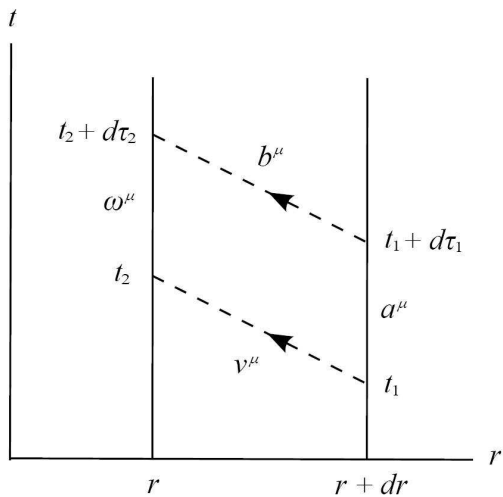}
\caption{The Pound-Rebka experiment}
\end{figure}

In Figure 1 the vectors $v^\mu$ and $b^\mu$ are null vectors that
represent the light signals. Null vectors satisfy the condition
$v^\mu v^\nu g_{\mu\nu}=0$. For radial null vectors we have
$v^0 v^0 g_{00}+v^1 v^1 g_{11}=0$, and therefore
$v^1=(-g_{00}/g_{11})^{1/2}\, v^0$. In the Schwarzschild spacetime
we have $v^1=(-g_{00}) v^0$. Thus a radial null vector in the 
Schwarzschild spacetime may be written as $v^\mu=v^0(1,-g_{00},0,0)$.

The vectors $a^\mu, b^\mu, v^\mu$ and $w^\mu$ in Figure 1 have 
dimension of length. Except for the factor $c$ (the speed of light), 
the zero components of these vectors represent the time elapsed
between two events. The time elapsed between $(t_1+d\tau_1, r+dr)$ 
and $(t_1, r+dr)$ is $d\tau_1=\beta(r+dr) dt$, and between
$(t_2+d\tau_2, r)$ and $(t_2, r)$ is $d\tau_2=\beta(r) dt$.
Thus, 

\begin{eqnarray}
a^\mu&=& \beta(r+dr)(cdt,0,0,0)\,, \nonumber \\
w^\mu&=& \beta(r) (cdt,0,0,0)\,.
\label{32}
\end{eqnarray}

Let us denote $dT$ the time elapsed between the events
$(t_2,r)$ and $(t_1, r+dr)$, or between
$(t_2+d\tau_2, r)$ and $(t_1+d\tau_1, r+dr)$. We write

\begin{eqnarray}
b^\mu&=& (c\,dT,-g_{00}\,c\,dT,0,0)\,, \nonumber \\
v^\mu&=& (c\,dT,-g_{00}\,c\,dT,0,0)\,.
\label{33}
\end{eqnarray}
However, ingoing radial null geodesics in the Schwarzschild spacetime
satisfy (see, for instance, section 16.4 of \cite{Inverno})

\begin{equation}
{{cdT}\over {dr}}=-{r\over {r-2m}}=-{1\over {1-2m/r}}=-g_{11}
={1\over {g_{00}}}\,.
\label{34}
\end{equation}
Thus, $c\,dT=(1/g_{00}) dr$, and finally we have

\begin{eqnarray}
b^\mu&=& dr({1\over {g_{00}}}, -1,0,0)\,, \nonumber \\
v^\mu&=& dr({1\over {g_{00}}}, -1,0,0) \,.
\label{35}
\end{eqnarray}

The breaking of the parallelogram in Figure 1 is verified by the
following operation,

\begin{eqnarray}
(a^\mu +b^\mu)-(v^\mu+w^\mu)&=&
\biggl( \beta(r+dr)-\beta(r)\biggr)(cdt,0,0,0)\,, \nonumber \\
&\cong& {m\over r^2} {{c\,dt\,dr}\over{(-g_{00})^{1/2}}}(1,0,0,0)\,.
\label{36}
\end{eqnarray}

The nonclosure of the parallelogram can also be obtained by
means of an alternative procedure. Let us consider two infinitesimal
vectors, $A^\mu=dx^\mu$ and $B^\mu=\delta x^\mu$, as in Figure 2 below.

\begin{figure}[h]
\centering
\includegraphics[width=0.60\textwidth]{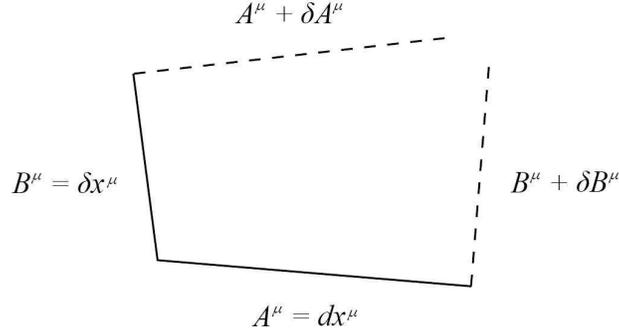}
\caption{The breaking of the parallelogram}
\end{figure}
The parallel transport of $A^\mu$ along $\delta x^\mu$, and of
$B^\mu$ along $dx^\mu$ are given by, respectively,

\begin{eqnarray}
\delta A^\mu&=& -\Gamma^\mu_{\alpha\beta} A^\alpha \delta x^\beta\,,
\nonumber \\
\delta B^\mu&=& -\Gamma^\mu_{\alpha\beta} B^\alpha dx^\beta\,,
\label{37}
\end{eqnarray}
where $\Gamma^\mu_{\alpha\beta}$ is a spacetime connection with 
torsion. The nonclosure of the parallelogram is obtained as follows,

\begin{eqnarray}
\lbrack A^\mu + (B^\mu +\delta B^\mu) \rbrack -
\lbrack B^\mu+ (A^\mu+\delta A^\mu)\rbrack 
&=&(\Gamma^\mu_{\alpha\beta}-
\Gamma^\mu_{\beta\alpha})dx^\alpha \delta x^\beta \nonumber \\
&=&T^\mu\,_{\alpha \beta}dx^\alpha \delta x^\beta\,.
\label{38}
\end{eqnarray}

As in the previous section, the Schwarzschild spacetime is described
by the set of tetrad fields given by (11), i.e., by stationary 
observers in spacetime. Without going into details of calculations
we just assert that the frame determined by (11) yields only three
components of the torsion tensor 
$T^\mu\,_{\alpha\beta}=e_a\,^\mu T^a\,_{\alpha\beta}$ (note that
there are six nonvanishing components of $T_{a\mu\nu}$),

\begin{eqnarray}
T^0\,_{01}&=& -{1\over \beta}\partial_1 \beta\,, \nonumber \\
T^2\,_{12}&=&T^3\,_{13}= {1\over r} (1-\alpha)\,.
\label{39}
\end{eqnarray}
Now we identify

\begin{eqnarray}
dx^\alpha&=& a^\alpha=\beta(r+dr)(c\,dt,0,0,0)\,, \nonumber \\
\delta x^\beta &=& v^\beta =dr({1\over {g_{00}}}, -1,0,0)\,.
\label{40}
\end{eqnarray}
It is straightforward to verify that 

\begin{eqnarray}
T^0\,_{\alpha\beta}dx^\alpha \delta x^\beta &=&
T^0\,_{01}a^0 v^1 \nonumber \\
&=&{m\over r^2} {1\over {(-g_{00})}} \beta(r+dr)\,c\,dt\,dr 
\nonumber \\
&\cong&{m\over r^2} {{c\,dt\,dr}\over{(-g_{00})^{1/2}}}\,.
\label{41}
\end{eqnarray}
As a consequence of eq. (39) no other breaking of parallelogram takes
place in the $(t,r)$ plane of Figure 1. 
Taking into account eq. (30), we may also write

\begin{equation}
T^0\,_{\alpha\beta}dx^\alpha \delta x^\beta=
{m\over r^2} {{c\,d\tau_1\,dr}\over{(-g_{00})}}\,.
\label{42}
\end{equation}

In view of the agreement between eqs. (36) and (41) we conclude that
the reference frame determined by (11) is indeed suitable to describe
the emergence of torsion in the Schwarzschild spacetime.

\section{Concluding remarks}

We have investigated two manifestations of torsion in the 
Schwarzschild spacetime in the framework of a set of tetrad fields
adapted to stationary observers. In order to maintain
a frame in stationary state in spacetime it is necessary to impart
to the frame a translational, inertial acceleration that exactly
cancels the gravitational, geodesic acceleration. By means of the
acceleration tensor defined by eq. (8), which is a coordinate
invariant definition, we have obtained the inertial acceleration and
consequently the geodesic acceleration on a radial trajectory. The
investigation of the expression of the geodesic acceleration led to
the conclusion that for certain values of $m$ and $r$ in the
Schwarzschild spacetime the curvature tensor approaches zero,
whereas the geodesic accceleration may acquire arbitrary values. We
concluded that it is the torsion tensor (of the Weitzenb\"ock
connection), and not the curvature tensor (of the Levi-Civita
connection) that is responsible for the geodesic acceleration of
a particle.

The same set of tetrad fields (eq. (11)) explains the Pound-Rebka
experiment in terms of breaking of parallelogram, as in Figure 1.
The set of tetrad fields given by (11) yields the torsion tensor
components (39), which are crucial to the agreement between (36)
(the nonclosure of the parallelogram directly from the 
Pound-Rebka experiment) and (41) (the breaking of the parallelogram
obtained by parallel transport).

We remark that the discussions and results of sections 3 and 4 are
valid for a wider class of spacetimes, namely, to all
spacetimes determined by the metric tensor

$$ds^2=-\beta^2dt^2 +\alpha^2 dr^2 +r^2d\theta^2+
r^2 \sin^2\theta d\phi^2\,,$$
for which $\beta(r)=1/\alpha(r)$ is an arbitrary function of the 
radial coordinate $r$. This class of spacetime metrics includes,
for instance, the Schwarzschild-de Sitter and the Reissner-Nordstrom
spacetimes. Let us briefly consider the de Sitter spacetime. We have
$\beta^2=1-Kr^2$, where $K$ is related to the positive cosmological
constant $\Lambda$ by means of $\Lambda =3K$. By repeating the
analysis of section 3 it follows from eqs. (15) and (16) that the
inertial acceleration that is necessary to impart to the frame such
that it remains stationary in spacetime (in the notation of eq. (18))
is given by

\begin{equation}
{\bf a}=(\partial_1 \beta)\hat {\bf r}=
-{{Kr}\over{(1-Kr^2)^{1/2}}}\hat {\bf r} \,.
\label{43}
\end{equation}
We see that when $r$ approaches the cosmological horizon 
$R=1/\sqrt{K}=\sqrt{3/\Lambda}$, ${\bf a}$ acquires arbitrarily large
values, whereas the curvature tensor component $R_{0101}$ remains 
finite and constant: $R_{0101}=-K$. Once again we see that the values 
of the acceleration tensor have no direct relationship to the 
curvature tensor.

The question finally arises: what is the connection of the 
Schwarzschild spacetime? Is it simply the frame independent, metric 
connection,

\begin{equation}
\Gamma^\lambda_{\mu\nu} =\,^0\Gamma^\lambda_{\mu\nu}\,,
\label{44}
\end{equation}
given by the Christoffel symbols, or a connection with torsion,

\begin{equation}
\Gamma^\lambda_{\mu\nu} =\,^0\Gamma^\lambda_{\mu\nu}+
T^\lambda\,_{\mu\nu}\,,
\label{45}
\end{equation}
where $T^\lambda\,_{\mu\nu}$ (obtained in the frame of stationary
observers) is given by (39)? Since 
$T^\lambda\,_{\mu\nu}$ is antisymmetric in the $\mu\nu$ indices, 
the geodesic equations obtained from connections (44) and (45) are
the same. $T^\lambda\,_{\mu\nu}$ does not affect the standard 
geodesic motion of particles in spacetime, and not any of the 
experimental tests of
general relativity. Moreover we note that in the limit $m/r<<1$
the three torsion components in eq. (39) fall off as $-m/r^2$.
It is likely that $T^\lambda\,_{\mu\nu}$ is relevant to
small scale gravitational phenomena.
However, (45) is not a metric compatible 
connection, as it leads to $\nabla_\alpha g_{\mu\nu} \ne 0$. 
The answer to the question above will probably require the
investigation of further experimental consequences of the 
Schwarzschild geometry.

\end{document}